\documentstyle[aps,graphicx]{revtex}

\def\beq{\begin{equation}}
\def\eeq{\end{equation}}
\def\beqa{\begin{eqnarray}}
\def\eeqa{\end{eqnarray}}

\def\za{\alpha}
\def\zb{\beta}
\def\lsim{\mathrel{\raise.3ex\hbox{$<$\kern-.75em\lower1ex\hbox{$\sim$}}} }
\def\gsim{\mathrel{\raise.3ex\hbox{$>$\kern-.75em\lower1ex\hbox{$\sim$}}} }

\begin{document}
\onecolumn

\begin{flushright}
NCU-HEP-k002\\
Aug 2001
\end{flushright}

\vspace*{1in}

\begin{center}
{\Large \bf Electric Dipole Moments in the Generic Supersymmetric Standard Model$^\star$}

\vspace*{.5in}
{\bf  Otto C.W. Kong}\\[.08in]
{\it Department of Physics, National Central University, Chung-li, TAIWAN 32054}\\[.05in]
{\it Institute of Physics, Academia Sinica, Nankang, Taipei, TAIWAN 11529}

\vspace*{1.5in}
{Abstract}\\
\end{center} 
The generic supersymmetric standard model is a model built from a 
supersymmetrized standard model field spectrum the gauge symmetries only. 
The popular minimal supersymmetric standard model differs from the 
generic version in having R-parity imposed by hand. We review an efficient 
formulation of the model and some of the recently obtained interesting 
phenomenological features, focusing on one-loop contributions to fermion 
electric dipole moments.

\noindent

\vfill
\noindent --------------- \\
$^\star$ Invited talk at NANP'01 conference (Jun 19 - 23), 
Dubna, Russia \\
 --- submission for the proceedings.  

\clearpage
\addtocounter{page}{-1}

\title{Electric Dipole Moments in the Generic Supersymmetric Standard Model}

\author{Otto C. W. Kong}

\address{Department of Physics, National Central University, Chung-li, TAIWAN 32054\footnote{Permanent address since Aug. 1, 2001.}\\
Institute of Physics, Academia Sinica, Nankang, Taipei, TAIWAN 11529
\\E-mail: otto@phy.ncu.edu.tw}
\maketitle

\begin{abstract}
The generic supersymmetric standard model is a model built from a 
supersymmetrized standard model field spectrum the gauge symmetries only. 
The popular minimal supersymmetric standard model differs from the 
generic version in having R-parity imposed by hand. We review an efficient 
formulation of the model and some of the recently obtained interesting 
phenomenological features, focusing on one-loop contributions to fermion 
electric dipole moments.
\end{abstract}

\section{Introduction}
Fermion electric dipole moments (EDMs) are known to be extremely useful 
constraints on (the CP violating part of) models depicting interesting 
scenarios of beyond Standard Model (SM) physics. In particular, the 
experimental bounds on neutron EDM ($d_n$) and electron EDM ($d_e$) are very 
stringent. The current numbers are given by 
$d_n < 6.3 \cdot 10^{-26}\,e \cdot \mbox{cm}$
and $d_e < 4.3 \cdot 10^{-27}\,e \cdot \mbox{cm}$. The SM contributions are 
known to be very small, given that the only source of CP violation has to 
come from the KM phase in (charged current) quark flavor mixings :
  $d_n \sim 10^{-32}\,e\cdot \mbox{cm}$ and 
$d_e \sim 8 \cdot 10^{-41}\,e\cdot \mbox{cm}$. 

Extensions of the SM normally are expected to have potentially large EDM 
contributions. For instance, for the minimal supersymmetric standard model (MSSM),
there are a few source of such new contributions. For example, they can come in 
through $LR$ sfermion mixings. The latter have two parts, an $A$-term contribution 
as well as a $F$-term contribution. The $F$-term is a result of the complex phase
in the so-called $\mu$-term. The resulted constraints on MSSM have been studied
extensively. We are interested here in the modified version with R parity 
not imposed. We will illustrate that there are extra contributions at the same 
level and discuss the class of important constraints hence resulted.\cite{me}

\section{The Generic Supersymmetric Standard Model}
A theory built with the minimal superfield spectrum incorporating the SM particles, 
the admissible renormalizable interactions dictated by the SM (gauge) symmetries 
together with the idea that supersymmetry (SUSY) is softly broken is what should 
be called the the generic supersymmetric standard model (GSSM). The popular
MSSM differs from the generic version in having a discrete symmetry, called R 
parity, imposed by hand to enforce baryon and lepton number conservation. With 
the strong experimental hints at the existence of lepton number violating neutrino 
masses, such a theory of SUSY without R-parity deserves ever more attention. The
GSSM contains all kinds of (so-called) R-parity violating (RPV) parameters.
The latter includes the more popular trilinear ($\lambda_{ijk}$, $\lambda_{ijk}^{\prime}$, and	$\lambda_{ijk}^{\prime\prime}$) and bilinear 
($\mu_i$) couplings in the superpotential, as well as  soft SUSY breaking
parameters of the trilinear, bilinear, and soft mass (mixing) types. In order not 
to miss any plausible RPV phenomenological features, it is important that all of 
the RPV parameters be taken into consideration without {\it a priori} bias. 
We do, however, expect some sort of symmetry principle to guard against the very 
dangerous proton decay problem. The emphasis is hence put on the lepton number
violating phenomenology.

The renormalizable superpotential for the GSSM can be written  as
\small\beqa
W \!\! &=& \!\varepsilon_{ab}\Big[ \mu_{\alpha}  \hat{H}_u^a \hat{L}_{\alpha}^b 
+ h_{ik}^u \hat{Q}_i^a   \hat{H}_{u}^b \hat{U}_k^{\scriptscriptstyle C}
+ \lambda_{\alpha jk}^{\!\prime}  \hat{L}_{\alpha}^a \hat{Q}_j^b
\hat{D}_k^{\scriptscriptstyle C} 
+
\frac{1}{2}\, \lambda_{\alpha \beta k}  \hat{L}_{\alpha}^a  
 \hat{L}_{\beta}^b \hat{E}_k^{\scriptscriptstyle C} \Big] + 
\frac{1}{2}\, \lambda_{ijk}^{\!\prime\prime}  
\hat{U}_i^{\scriptscriptstyle C} \hat{D}_j^{\scriptscriptstyle C}  
\hat{D}_k^{\scriptscriptstyle C}   ,
\eeqa\normalsize
where  $(a,b)$ are $SU(2)$ indices, $(i,j,k)$ are the usual family (flavor) 
indices, and $(\za, \zb)$ are extended flavor indices going from $0$ to $3$.
At the limit where $\lambda_{ijk}, \lambda^{\!\prime}_{ijk},  
\lambda^{\!\prime\prime}_{ijk}$ and $\mu_{i}$  all vanish, 
one recovers the expression for the R-parity preserving MSSM, 
with $\hat{L}_{0}$ identified as $\hat{H}_d$. Without R-parity imposed,
the latter is not {\it a priori} distinguishable from the $\hat{L}_{i}$'s.
Note that $\lambda$ is antisymmetric in the first two indices, as
required by  the $SU(2)$  product rules, as shown explicitly here with 
$\varepsilon_{\scriptscriptstyle 12} =-\varepsilon_{\scriptscriptstyle 21}=1$.
Similarly, $\lambda^{\!\prime\prime}$ is antisymmetric in the last two 
indices, from $SU(3)_{\scriptscriptstyle C}$. 

R-parity is exactly an {\it ad hoc} symmetry put in to make $\hat{L}_{0}$,
stand out from the other $\hat{L}_i$'s as the candidate for  $\hat{H}_d$.
It is defined in terms of baryon number, lepton number, and spin as, 
explicitly, ${\mathcal R} = (-1)^{3B+L+2S}$. The consequence is that 
the accidental symmetries of baryon number and lepton number in the SM 
are preserved, at the expense of making particles and superparticles having 
a categorically different quantum number, R parity. The latter is actually 
not the most effective discrete symmetry to control superparticle 
mediated proton decay\cite{pd}, but is most restrictive in terms
of what is admitted in the Lagrangian, or the superpotential alone. 
On the other hand, R parity also forbides neutrino masses in the
supersymmetric SM. The strong experimental hints for the existence of 
(Majorana) neutrino masses\cite{exp} is an indication of lepton 
number violation, hence suggestive of R-parity violation.

The soft SUSY breaking part 
of the Lagrangian is more interesting, if only for the fact that  many
of its interesting details have been overlooked in the literature.
However, we will postpone the discussion till after we address the
parametrization issue.

\section{Parametrization}
Doing phenomenological studies without specifying a choice 
of flavor bases is ambiguous. It is like doing SM quark physics with 18
complex Yukawa couplings, instead of the 10 real physical parameters.
As far as the SM itself is concerned, the extra 26 real parameters
are simply redundant, and attempts to relate the full 36 parameters to
experimental data will be futile. In the GSSM, the choice of an optimal
parametrization mainly concerns the 4 $\hat{L}_\alpha$ flavors. We use
here the single-VEV parametrization\cite{ru,as8} (SVP), in which flavor 
bases are chosen such that : 
1/ among the $\hat{L}_\alpha$'s, only  $\hat{L}_0$, bears a VEV,
{\it i.e.} {\small $\langle \hat{L}_i \rangle \equiv 0$};
2/  {\small $h^{e}_{jk} (\equiv \lambda_{0jk}) 
=\frac{\sqrt{2}}{v_{\scriptscriptstyle 0}} \,{\rm diag}
\{m_{\scriptscriptstyle 1},
m_{\scriptscriptstyle 2},m_{\scriptscriptstyle 3}\}$};
3/ {\small $h^{d}_{jk} (\equiv \lambda^{\!\prime}_{0jk} =-\lambda_{j0k}) 
= \frac{\sqrt{2}}{v_{\scriptscriptstyle 0}}{\rm diag}\{m_d,m_s,m_b\}$}; 
4/ {\small $h^{u}_{ik}=\frac{\sqrt{2}}{v_{\scriptscriptstyle u}}
V_{\mbox{\tiny CKM}}^{\!\scriptscriptstyle T} \,{\rm diag}\{m_u,m_c,m_t\}$}, 
where ${v_{\scriptscriptstyle 0}} \equiv  \sqrt{2}\,\langle \hat{L}_0 \rangle$
and ${v_{\scriptscriptstyle u} } \equiv \sqrt{2}\,
\langle \hat{H}_{u} \rangle$. The big advantage of the SVP is that it gives 
the complete tree-level mass matrices of all the states (scalars and fermions) 
the simplest structure\cite{as5,as8}.

\section{Leptons in GSSM}
The SVP gives quark mass matrices exactly in the SM form. For the masses
of the color-singlet fermions, all the RPV effects are paramatrized by the
$\mu_i$'s only. For example, the five charged fermions ( gaugino
+ Higgsino + 3 charged leptons ), we have
\small\beq \label{mc}
{\mathcal{M}_{\scriptscriptstyle C}} =
 \left(
{\begin{array}{ccccc}
{M_{\scriptscriptstyle 2}} &  
\frac{g_{\scriptscriptstyle 2}{v}_{\scriptscriptstyle 0}}{\sqrt 2}  
& 0 & 0 & 0 \\
 \frac{g_{\scriptscriptstyle 2}{v}_{\scriptscriptstyle u}}{\sqrt 2} & 
 {{ \mu}_{\scriptscriptstyle 0}} & {{ \mu}_{\scriptscriptstyle 1}} &
{{ \mu}_{\scriptscriptstyle 2}}  & {{ \mu}_{\scriptscriptstyle 3}} \\
0 &  0 & {{m}_{\scriptscriptstyle 1}} & 0 & 0 \\
0 & 0 & 0 & {{m}_{\scriptscriptstyle 2}} & 0 \\
0 & 0 & 0 & 0 & {{m}_{\scriptscriptstyle 3}}
\end{array}}
\right)  \; .
\eeq\normalsize
Moreover each $\mu_i$ parameter here characterizes directly the RPV effect
on the corresponding charged lepton  ($\ell_i = e$, $\mu$, and $\tau$).
This, and the corresponding neutrino-neutralino masses and mixings,
has been exploited to implement a detailed study of the tree-level
RPV phenomenology from the gauge interactions, with interesting 
results\cite{ru}.

Neutrino masses and oscillations is no doubt one of the most important aspects
of the model. Here, it is particularly important that the various RPV 
contributions to neutrino masses, up to 1-loop level, be studied in a 
framework that takes no assumption on the other parameters. Our formulation 
provides such a framework. Interested readers are referred to 
Refs.\cite{ok,as1,as5,as9,AL}.

\section{Soft SUSY Breaking Terms and the Scalar Masses}
Obtaining the squark and slepton masses is straightforward, once all the 
admissible soft SUSY breaking terms are explicitly written down\cite{as5}. 
The soft SUSY breaking part of the Lagrangian can be written as 
\beqa
V_{\rm soft} &=& \epsilon_{\!\scriptscriptstyle ab} 
  B_{\za} \,  H_{u}^a \tilde{L}_\za^b +
\epsilon_{\!\scriptscriptstyle ab} \left[ \,
A^{\!\scriptscriptstyle U}_{ij} \, 
\tilde{Q}^a_i H_{u}^b \tilde{U}^{\scriptscriptstyle C}_j 
+ A^{\!\scriptscriptstyle D}_{ij} 
H_{d}^a \tilde{Q}^b_i \tilde{D}^{\scriptscriptstyle C}_j  
+ A^{\!\scriptscriptstyle E}_{ij} 
H_{d}^a \tilde{L}^b_i \tilde{E}^{\scriptscriptstyle C}_j   \,
\right] + {\rm h.c.}\nonumber \\
&+&
\epsilon_{\!\scriptscriptstyle ab} 
\left[ \,  A^{\!\scriptscriptstyle \lambda^\prime}_{ijk} 
\tilde{L}_i^a \tilde{Q}^b_j \tilde{D}^{\scriptscriptstyle C}_k  
+ \frac{1}{2}\, A^{\!\scriptscriptstyle \lambda}_{ijk} 
\tilde{L}_i^a \tilde{L}^b_j \tilde{E}^{\scriptscriptstyle C}_k  
\right] 
+ \frac{1}{2}\, A^{\!\scriptscriptstyle \lambda^{\prime\prime}}_{ijk} 
\tilde{U}^{\scriptscriptstyle C}_i  \tilde{D}^{\scriptscriptstyle C}_j  
\tilde{D}^{\scriptscriptstyle C}_k  + {\rm h.c.}
\nonumber \\
&+&
 \tilde{Q}^\dagger \tilde{m}_{\!\scriptscriptstyle {Q}}^2 \,\tilde{Q} 
+\tilde{U}^{\dagger} 
\tilde{m}_{\!\scriptscriptstyle {U}}^2 \, \tilde{U} 
+\tilde{D}^{\dagger} \tilde{m}_{\!\scriptscriptstyle {D}}^2 
\, \tilde{D} 
+ \tilde{L}^\dagger \tilde{m}_{\!\scriptscriptstyle {L}}^2  \tilde{L}  
  +\tilde{E}^{\dagger} \tilde{m}_{\!\scriptscriptstyle {E}}^2 
\, \tilde{E}
+ \tilde{m}_{\!\scriptscriptstyle H_{\!\scriptscriptstyle u}}^2 \,
|H_{u}|^2 
\nonumber \\
&& + \frac{M_{\!\scriptscriptstyle 1}}{2} \tilde{B}\tilde{B}
   + \frac{M_{\!\scriptscriptstyle 2}}{2} \tilde{W}\tilde{W}
   + \frac{M_{\!\scriptscriptstyle 3}}{2} \tilde{g}\tilde{g}
+ {\rm h.c.}\; ,
\label{soft}
\eeqa
where we have separated the R-parity conserving $A$-terms from the 
RPV ones (recall $\hat{H}_{d} \equiv \hat{L}_0$). Note that 
$\tilde{L}^\dagger \tilde{m}_{\!\scriptscriptstyle \tilde{L}}^2  \tilde{L}$,
unlike the other soft mass terms, is given by a 
$4\times 4$ matrix. Explicitly, 
$\tilde{m}_{\!\scriptscriptstyle {L}_{00}}^2$ corresponds to 
$\tilde{m}_{\!\scriptscriptstyle H_{\!\scriptscriptstyle d}}^2$ 
of the MSSM case while 
$\tilde{m}_{\!\scriptscriptstyle {L}_{0k}}^2$'s give RPV mass mixings.

The only RPV contribution to the squark masses is given by a
$- (\, \mu_i^*\lambda^{\!\prime}_{ijk}\, ) \; 
\frac{v_{\scriptscriptstyle u}}{\sqrt{2}}$ term in the $LR$ mixing part.
Note that the term contains flavor-changing ($j\ne k$) parts which,
unlike the $A$-terms ones, cannot be suppressed through a flavor-blind
SUSY breaking spectrum. Hence, it has very interesting implications
to quark electric dipole moments (EDMs) and related processses
such as $b\to s\, \gamma$\cite{as4,as6,kk,cch1}.

The mass matrices are a bit more complicated in the scalar sectors\cite{as5,as7}.
The $1+4+3$ charged scalar masses are given in terms of the blocks
\small\beqa
&& \widetilde{\cal M}_{\!\scriptscriptstyle H\!u}^2 =
\tilde{m}_{\!\scriptscriptstyle H_{\!\scriptscriptstyle u}}^2
+ \mu_{\!\scriptscriptstyle \za}^* \mu_{\scriptscriptstyle \za}^{}
+ M_{\!\scriptscriptstyle Z}^2\, \cos\!2 \beta 
\left[ \,\frac{1}{2} - \sin\!^2\theta_{\!\scriptscriptstyle W}\right]
+ M_{\!\scriptscriptstyle Z}^2\,  \sin\!^2 \beta \;
[1 - \sin\!^2 \theta_{\!\scriptscriptstyle W}]
\; ,
\nonumber \\
&&\widetilde{\cal M}_{\!\scriptscriptstyle LL}^2
= \tilde{m}_{\!\scriptscriptstyle {L}}^2 +
m_{\!\scriptscriptstyle L}^\dag m_{\!\scriptscriptstyle L}^{}
+ M_{\!\scriptscriptstyle Z}^2\, \cos\!2 \beta 
\left[ -\frac{1}{2} +  \sin\!^2 \theta_{\!\scriptscriptstyle W}\right] 
+ \left( \begin{array}{cc}
 M_{\!\scriptscriptstyle Z}^2\,  \cos\!^2 \beta \;
[1 - \sin\!^2 \theta_{\!\scriptscriptstyle W}] 
& \quad 0_{\scriptscriptstyle 1 \times 3} \quad \\
0_{\scriptscriptstyle 3 \times 1} & 0_{\scriptscriptstyle 3 \times 3}  
\end{array} \right) 
+ (\mu_{\!\scriptscriptstyle \za}^* \mu_{\scriptscriptstyle \zb}^{})
\; ,
\nonumber \\
&& \widetilde{\cal M}_{\!\scriptscriptstyle RR}^2 =
\tilde{m}_{\!\scriptscriptstyle {E}}^2 +
m_{\!\scriptscriptstyle E}^{} m_{\!\scriptscriptstyle E}^\dag
+ M_{\!\scriptscriptstyle Z}^2\, \cos\!2 \beta 
\left[  - \sin\!^2 \theta_{\!\scriptscriptstyle W}\right] \; ; \qquad
\eeqa
{\normalsize and}
\beqa 
\label{ELH}
\widetilde{\cal M}_{\!\scriptscriptstyle LH}^2
&=& (B_{\za}^*)  
+ \left( \begin{array}{c} 
{1 \over 2} \,
M_{\!\scriptscriptstyle Z}^2\,  \sin\!2 \beta \,
[1 - \sin\!^2 \theta_{\!\scriptscriptstyle W}]  \\
0_{\scriptscriptstyle 3 \times 1} 
\end{array} \right)\; ,
\qquad
\\
\label{ERH}
\widetilde{\cal M}_{\!\scriptscriptstyle RH}^2
&=&  -\,(\, \mu_i^*\lambda_{i{\scriptscriptstyle 0}k}\, ) \; 
\frac{v_{\scriptscriptstyle 0}}{\sqrt{2}} \; ,
\\ 
\label{ERL}
(\widetilde{\cal M}_{\!\scriptscriptstyle RL}^{2})^{\scriptscriptstyle T} 
&=& \left(\begin{array}{c} 
0  \\   A^{\!{\scriptscriptstyle E}} 
\end{array}\right)
 \frac{v_{\scriptscriptstyle 0}}{\sqrt{2}}
-\,(\, \mu_{\scriptscriptstyle \za}^*
\lambda_{{\scriptscriptstyle \za\zb}k}\, ) \, 
\frac{v_{\scriptscriptstyle u}}{\sqrt{2}} \; .
\eeqa \normalsize
For the neutral scalars, we have explicitly
\beq \label{MSN}
{\cal M}_{\!\scriptscriptstyle S}^2 =
\left( \begin{array}{cc}
{\cal M}_{\!\scriptscriptstyle SS}^2 &
{\cal M}_{\!\scriptscriptstyle SP}^2 \\
({\cal M}_{\!\scriptscriptstyle SP}^{2})^{\!\scriptscriptstyle T} &
{\cal M}_{\!\scriptscriptstyle PP}^2
\end{array} \right) \; ,
\eeq
where the scalar, pseudo-scalar, and mixing parts are given by
\beqa
{\cal M}_{\!\scriptscriptstyle SS}^2 &=&
\mbox{Re}({\cal M}_{\!\scriptscriptstyle {\phi}{\phi}\dag}^2)
+ {\cal M}_{\!\scriptscriptstyle {\phi\phi}}^2 \; ,
\nonumber \\
{\cal M}_{\!\scriptscriptstyle PP}^2 &=&
\mbox{Re}({\cal M}_{\!\scriptscriptstyle {\phi}{\phi}\dag}^2)
- {\cal M}_{\!\scriptscriptstyle {\phi\phi}}^2 \; ,
\nonumber \\
{\cal M}_{\!\scriptscriptstyle SP}^2 &=& -
\mbox{Im}({\cal M}_{\!\scriptscriptstyle {\phi}{\phi}\dag}^2) \; ,
\label{lastsc}
\eeqa
respectively
\footnote{Note that the original expression given in Ref.\cite{as5} has a 
typo in the  ${\cal M}_{\!\scriptscriptstyle SP}^2$ expression.}, with 
\small\beqa \label{Mpp}
{\cal M}_{\!\scriptscriptstyle {\phi\phi}}^2 
 &=& {1\over 2} \, M_{\!\scriptscriptstyle Z}^2\,
\left( \begin{array}{ccc}
 \sin\!^2\! \beta   &  - \cos\!\beta \, \sin\! \beta
 &  \quad 0_{\scriptscriptstyle 1 \times 3} \\
 - \cos\!\beta \, \sin\! \beta \!\! & \!\! \cos\!^2\! \beta 
 &  \quad 0_{\scriptscriptstyle 1 \times 3} \\
0_{\scriptscriptstyle 3 \times 1} &  0_{\scriptscriptstyle 3 \times 1} 
 & \quad 0_{\scriptscriptstyle 3 \times 3} 
\end{array} \right) ,
\eeqa \normalsize
{\normalsize and}
\beqa 
{\cal M}_{\!\scriptscriptstyle {\phi}{\phi}^\dag}^2 
 &=&  {\cal M}_{\!\scriptscriptstyle {\phi\phi}}^2 +
 \left(  \begin{array}{cc}
\tilde{m}_{\!\scriptscriptstyle H_{\!\scriptscriptstyle u}}^2
+ \mu_{\!\scriptscriptstyle \za}^* \mu_{\scriptscriptstyle \za}
  -\frac{1}{2} \, M_{\!\scriptscriptstyle Z}^2\, \cos\!2 \beta
& - (B_\za) \\
- (B_\za^*) &
\tilde{m}_{\!\scriptscriptstyle {L}}^2 
+ (\mu_{\!\scriptscriptstyle \za}^* \mu_{\scriptscriptstyle \zb})
+ \frac{1}{2} \, M_{\!\scriptscriptstyle Z}^2\, \cos\!2 \beta
\end{array}  \right) \; .
\label{Mp}
\eeqa \normalsize
Note that $\tilde{m}_{\!\scriptscriptstyle {L}}^2$ here is a $4\times 4$
matrix of soft masses for the $L_\za$, and $B_\za$'s are the corresponding 
bilinear soft terms of the $\mu_{\scriptscriptstyle \za}$'s.
$A^{\!{\scriptscriptstyle E}}$ is just the $3\times 3$ R-parity conserving
leptonic $A$-term. There is no contribution from the admissible RPV $A$-terms
under the SVP. Also, we have used
$m_{\!\scriptscriptstyle L} \equiv \mbox{diag} \{\,0, m_{\!\scriptscriptstyle E}\,\}
\equiv \mbox{diag} \{\,0, m_{\scriptscriptstyle 1}, m_{\scriptscriptstyle 2}, 
m_{\scriptscriptstyle 3}\,\}$.

\section{Neutron Electric Dipole Moment}
Let us take a look first at the quark dipole operator through 1-loop diagrams
with $LR$ squark mixing. A simple direct example is given by the gluino diagram.
Comparing with the MSSM case, the extra (RPV) to the $d$ squark $LR$ mixing
in GSSM obvious modified the story. If
one naively imposes the constraint for this RPV contribution 
itself not to exceed the experimental bound on neutron EDM, one gets roughly
$\mbox{Im}(\mu_i^*\lambda^{\!\prime}_{i\scriptscriptstyle 1\!1}) 
\lsim 10^{-6}\,\mbox{GeV}$, a constraint that is interesting even
in comparison to the bounds on the corresponding parameters obtainable
from asking no neutrino masses to exceed the super-Kamiokande 
atmospheric oscillation scale\cite{as4}. 

In fact, there are important contributions beyond the gluino diagram and without 
$LR$ squark mixings involved. For the MSSM, it is well-known that there is such
a contribution from the chargino diagram, which is likely to be more important
than the gluino one when a unification type gaugino mass relationship is
imposed. The question then is if the GSSM has a similar RPV analog. A RPV
version of the chargino diagram is given in Fig.1. The diagram, however, looks
ambiguous. Looking at the diagram in terms of the electroweak states involved
under our formulation, it seems like a 
${l}_k^{\!\!\mbox{ -}}$--$\tilde{W}^{\scriptscriptstyle +}$
mass insertion is required, which is however vanishing. However, putting in
extra mass insertion, with a $\mu_i$ flipping the ${l}_k^{\!\!\mbox{ -}}$ into 
a $\tilde{h}_u^{\scriptscriptstyle +}$ first seems to give a non-zero result.
The structure obviously indicates a GIM-like cancellation at worked, and we
have to check its violation due to the lack of mass degeneracy.

\begin{figure}[h]
\includegraphics{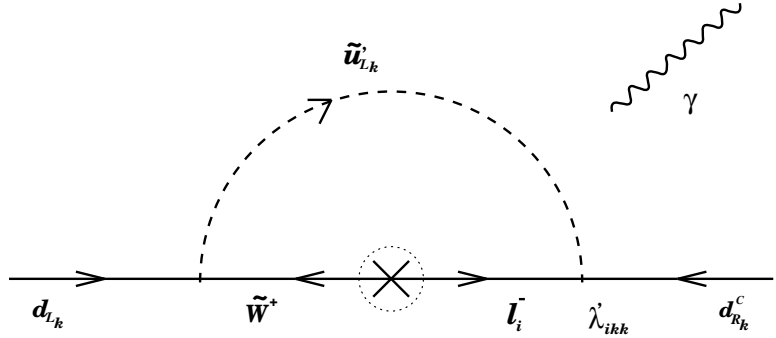}
\vspace*{2in}
\caption{The new charginolike diagram.}
\end{figure}
\vspace*{.5in}

We have performed an extensive analytical and numerical study, including the 
complete charginolike contributions, as well as the neutralinolike contributions,
to the neutron EDM\cite{as6}. The charginolike part is given by the following
formula :
\beq \label{edmco}
\left({d_{\scriptscriptstyle f} \over e} \right)_{\!\!\chi^{\mbox{-}}} = 
{\alpha_{\!\mbox{\tiny em}} \over 4 \pi \,\sin\!^2\theta_{\!\scriptscriptstyle W}} \; 
\sum_{\scriptscriptstyle \tilde{f}'\mp} 
\sum_{n=1}^{5} \,\mbox{Im}({\cal C}_{\!fn\mp}) \;
{{M}_{\!\scriptscriptstyle \chi^{\mbox{-}}_n} \over 
M_{\!\scriptscriptstyle \tilde{f}'\mp}^2} \;
\left[ {\cal Q}_{\!\tilde{f}'} \; 
B\!\left({{M}_{\!\scriptscriptstyle \chi^{\mbox{-}}_{n}}^2 \over 
M_{\!\scriptscriptstyle \tilde{f}'\mp}^2} \right) 
+ ( {\cal Q}_{\!{f}} - {\cal Q}_{\!\tilde{f}'} ) \;
A\!\left({{M}_{\!\scriptscriptstyle \chi^{\mbox{-}}_{n}}^2 \over 
M_{\!\scriptscriptstyle \tilde{f}'\mp}^2} \right) 
\right] \; ,
\eeq 
for $f$ being $u$ ($d$) quark and $f'$ being $d$ ($u$), where
\beqa
{\cal C}_{un-} &=&  
{y_{\!\scriptscriptstyle u} \over g_{\scriptscriptstyle 2} } \,\, 
\mbox{\boldmath $V$}^{\!*}_{\!\!2n} \, {\cal D}_{d11} \;
\left( - \mbox{\boldmath $U$}_{\!1n} \,{\cal D}^{*}_{d11} 
+ {y_{\!\scriptscriptstyle d} \over g_{\scriptscriptstyle 2} }\,\, 
\mbox{\boldmath $U$}_{\!2n}\,  {\cal D}^{*}_{d21}
+ {\lambda^{\!\prime}_{k\scriptscriptstyle 1\!1} \over g_{\scriptscriptstyle 2} }\,\, 
\mbox{\boldmath $U$}_{\!(k+2)n}\,  {\cal D}^{*}_{d21} \right) \; ,
\nonumber \\
{\cal C}_{un+} &=&
 {y_{\!\scriptscriptstyle u} \over g_{\scriptscriptstyle 2} } \,\, 
\mbox{\boldmath $V$}^{\!*}_{\!\!2n} \, {\cal D}_{d12} \;
\left( - \mbox{\boldmath $U$}_{\!1n} \, {\cal D}^{*}_{d12} 
+ {y_{\!\scriptscriptstyle d} \over g_{\scriptscriptstyle 2} }\,\, 
\mbox{\boldmath $U$}_{\!2n}\,  {\cal D}^{*}_{d22}
+ {\lambda^{\!\prime}_{k\scriptscriptstyle 1\!1} \over g_{\scriptscriptstyle 2} }\,\, 
\mbox{\boldmath $U$}_{\!(k+2)n}\,  {\cal D}^{*}_{d22} \right) \; ,
\nonumber \\
{\cal C}_{dn-} &=& 
\left( {y_{\!\scriptscriptstyle d} \over g_{\scriptscriptstyle 2} }\,\, 
\mbox{\boldmath $U$}_{\!2n} 
+ {\lambda^{\!\prime}_{k\scriptscriptstyle 1\!1} \over g_{\scriptscriptstyle 2} }\,\, 
\mbox{\boldmath $U$}_{\!(k+2)n} \right)\! {\cal D}_{u11} \;
\left( - \mbox{\boldmath $V$}^{\!*}_{\!\!1n} \,{\cal D}^{*}_{u11} 
+ {y_{\!\scriptscriptstyle u} \over g_{\scriptscriptstyle 2} } \,
\mbox{\boldmath $V$}^{\!*}_{\!\!2n} \, {\cal D}^{*}_{u21} \right) \; ,
\nonumber \\
{\cal C}_{dn+} &=& 
\left( {y_{\!\scriptscriptstyle d} \over g_{\scriptscriptstyle 2} }\,\, 
\mbox{\boldmath $U$}_{\!2n} 
+ {\lambda^{\!\prime}_{k\scriptscriptstyle 1\!1} \over g_{\scriptscriptstyle 2} }\,\, 
\mbox{\boldmath $U$}_{\!(k+2)n} \right)\! {\cal D}_{u12} \;
\left( - \mbox{\boldmath $V$}^{\!*}_{\!\!1n} \, {\cal D}^{*}_{u12} 
+ {y_{\!\scriptscriptstyle u} \over g_{\scriptscriptstyle 2} } \,
 \mbox{\boldmath $V$}^{\!*}_{\!\!2n} \, {\cal D}^{*}_{u22} \right) \; ,
\nonumber \\
&& \mbox{\hspace*{2.5in}\small(only repeated index $i$ is to be summed)} \; ;
\label{Cnmp}
\eeqa
$\mbox{\boldmath $V$}^\dag {\mathcal{M}_{\scriptscriptstyle C}} \,
\mbox{\boldmath $U$} = \mbox{diag} 
\{ {M}_{\!\scriptscriptstyle \chi^{\mbox{-}}_n} \} \equiv 
\mbox{diag} 
\{ {M}_{c {\scriptscriptstyle 1}}, {M}_{c {\scriptscriptstyle 2}},
m_e, m_\mu, m_\tau \}$ while ${\cal D}_{u}$ and ${\cal D}_{d}$ diagonalize
the $\tilde{u}$ and $\tilde{d}$ squark mass-squared matrices respectively;
and
\beq
A(x) = {1 \over 2 \, (1-x)^2} \left(3 - x + {2\ln x \over 1-x} \right)\;,
\qquad \qquad
B(x) = {1 \over 2\,(x-1)^2} \left[1 + x + {2\,x \ln x \over (1-x) } \right]. 
\nonumber 
\eeq

To extract the contribution from the diagram of Fig.~1, we have to look at the
pieces in ${\cal C}_{dn\mp}$ with a $\mbox{\boldmath $V$}^{\!*}_{\!\!1n}$
and a $\mbox{\boldmath $U$}_{\!(k+2)n}$. It is easy to see that the $n=1$ and 
$2$ mass eigenstates, namely the chargino states, do give the dominating
contribution. With the small $\mu_i$ mixings strongly favored by the 
sub-eV neutrino masses, we have 
\beq
\mbox{\boldmath $U$}_{\!(k+2)1} =
\frac{{\mu_k^*}}{{M}_{c {\scriptscriptstyle 1}}}
R_{\!\scriptscriptstyle R_{21}}  
\qquad \mbox{and} \qquad
\mbox{\boldmath $U$}_{\!(k+2)2} =
\frac{{\mu_k^*}}{{M}_{c {\scriptscriptstyle 2}}}
R_{\!\scriptscriptstyle R_{22}} 
\eeq
where the $R_{\!\scriptscriptstyle R}$ denotes the right-handed rotation 
that would diagonalize the first $2\times 2$ block of 
${\mathcal{M}_{\scriptscriptstyle C}}$. The latter rotation matrix is expected
to have elements of order 1. Hence, we have the dominating result
proportional to 
\[ 
\sum_{n=1,2}
R_{\!\scriptscriptstyle R_{12}}^{\,*} \,
R_{\!\scriptscriptstyle R_{2}n} \;
{\mu_k^* \,\lambda^{\!\prime}_{k\scriptscriptstyle 1\!1}} \;
F_{\!\scriptscriptstyle B\!A}\!\!
\left( M_{c_{\scriptscriptstyle n}}^2 \right) 
\]
where $F_{\!\scriptscriptstyle B\!A}$ denotes the mass eigenvalue
dependent part. The result agrees with what we say above. It vanishes for
${M}_{c {\scriptscriptstyle 1}}={M}_{c {\scriptscriptstyle 2}}$, showing
a GIM-like mechanism. However, with unequal chargino masses, our numerical
results indicate that the cancellation is generically badly violated.
More interestingly, it can be seen from the above analysis that a complex
phase in ${\mu_k^* \,\lambda^{\!\prime}_{k\scriptscriptstyle 1\!1}}$
is actually no necessary for this potentially dominating chargino 
contribution to be there, so long as complex CP violating phases exist
in the $R_{\!\scriptscriptstyle R}$ matrix, {\it i.e.} in the R-parity
conserving parameters such as ${\mu_{\scriptscriptstyle 0}}$.

\begin{center}{\rule{5in}{.01mm}}\end{center}

\begin{minipage}[b]{\textwidth}
\twocolumn

\begin{figure}[t]
\vspace*{4in}
\includegraphics{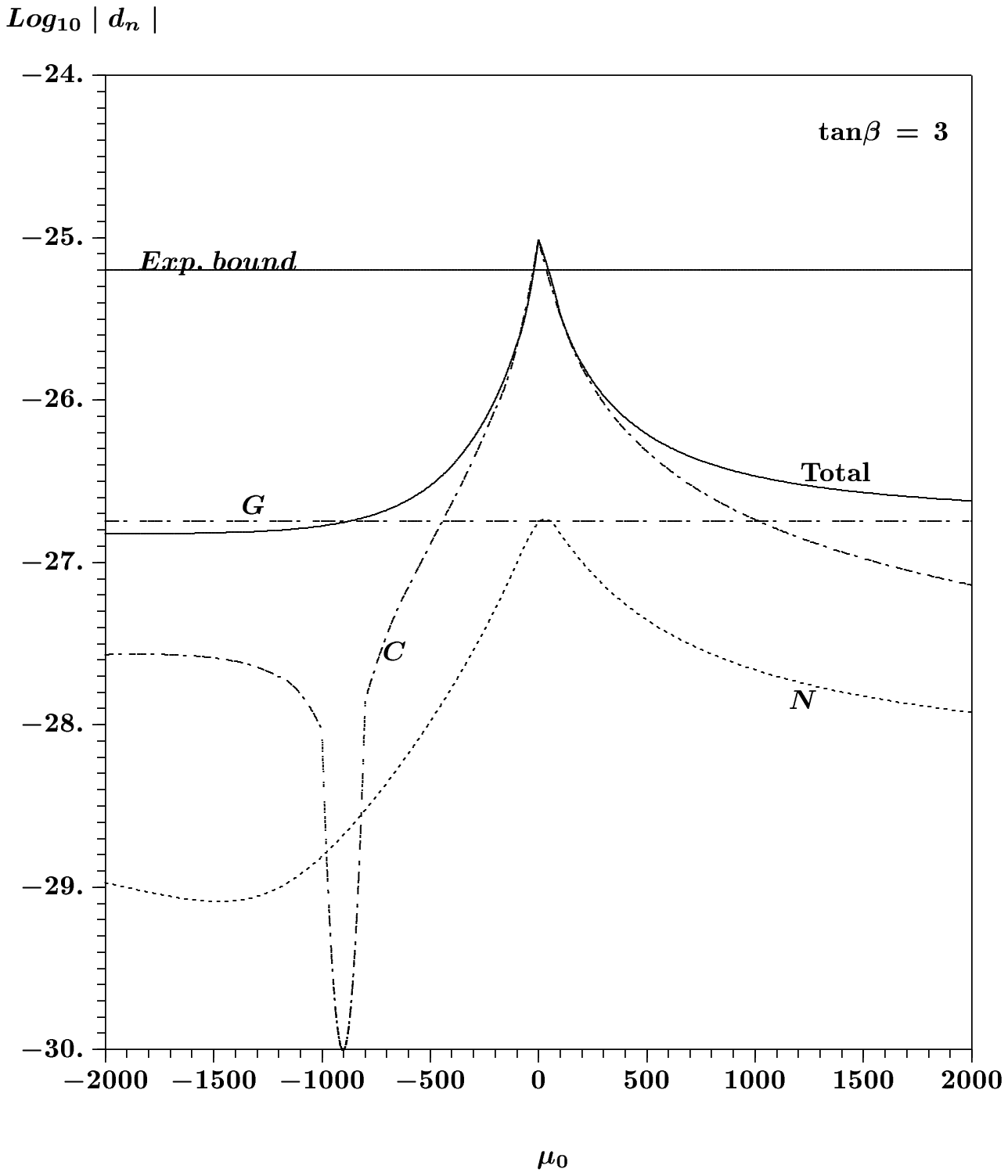}
\end{figure}
\parbox{3.2in}{\small FIG.2
Logarithmic plot of (the magnitude of) the RPV neutron EDM 
result for $\mu_{\!\scriptscriptstyle 0}$ value between $\pm2000\,\mbox{GeV}$,  
with the other parameters set at the same values as Case~A in Table~I. 
The lines marked by $G$, $C$, $N$, and ``Total" gives the complete
gluino, chargino-like, neutralino-like, and total ({\it i.e.} sum of the three) contributions, respectively. Note that the values of the $N$
contributions and those of the $C$ line for 
$\mu_{\!\scriptscriptstyle 0}<-900\,\mbox{GeV}$ are negative. 
}
\end{minipage}

\begin{minipage}[b]{\textwidth}
\vspace*{3.2in}
\begin{figure}[b]
\vspace*{4in}
\includegraphics{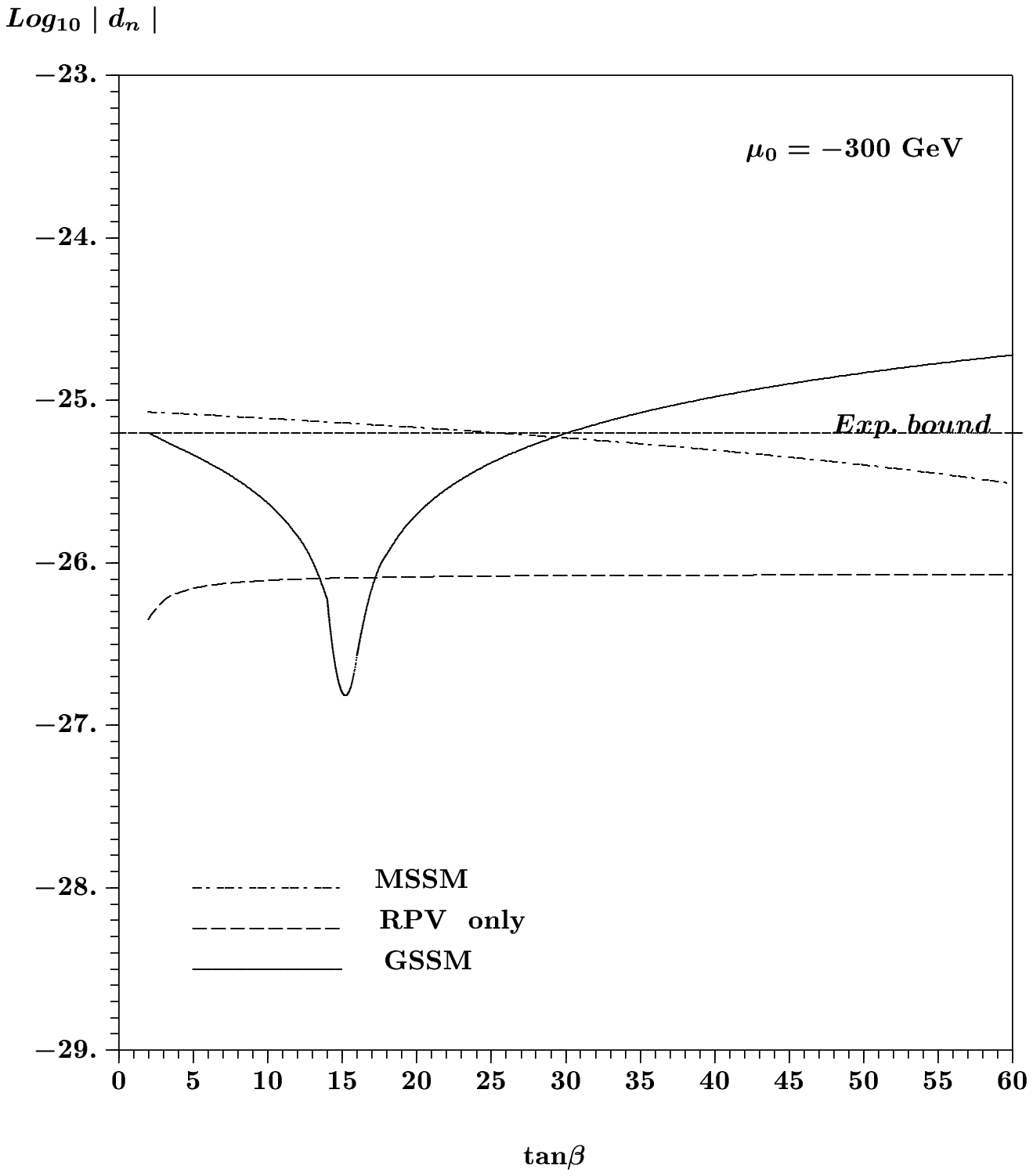}
\end{figure}
\vspace*{-.1in}
\parbox{3.2in}{\small FIG.3
Logarithmic plot of (the magnitude of) the neutron EDM result verses 
$\tan\!\zb$. We show here the MSSM result, our general result with RPV phase
only, and the generic result  with complex phases of 
both kinds. In particular, the $A$ and $\mu_{\scriptscriptstyle 0}$ 
phases are chosen as $7^o$ and $0.1^o$ respectively, for the MSSM line. They are 
zero for the RPV-only line, with which we have a phase of ${\pi\over 4}$ for 
$\lambda^{\!\prime}_{\scriptscriptstyle 31\!1}$. All the given nonzero values
are used for the three phases for the generic result (from our complete formulae) 
marked by GSSM. Again, the other unspecified input parameters are the same as for 
Case~A of Table~I.
\vspace*{.5in}}
\end{minipage}

\onecolumn

Some ilustarative sets of our numerical results are presented in Table~I.
In Fig.~2, we illustrate a comparison of the gluino, charginolike, and
neutralinolike contributions for a range of ${\mu_{\scriptscriptstyle 0}}$
values. Fig.~3 gives variation against the $\tan\!\zb$ value, while comparing
the overall GSSM result with the MSSM result.
On the whole, the magnitude of the parameter combination
$\mu_i^*\lambda^{\!\prime}_{i\scriptscriptstyle 1\!1}$ is shown to be 
responsible for the RPV 1-loop contribution to neutron EDM and is hence 
well constrained. This applies not only to the complex phase, or imaginary part 
of, the combination. 

\section{EDMs of the Electron and Other Fermions}
The above quark EDM formula obviously applies with some trivial modifications
to the cases of the other quarks. For the leptons, while the exact formulae
would be different, there are major basic features that are more or less
the same. For instance, for the charged lepton, the $\lambda$-couplings
play the role of the $\lambda^{\!\prime}$-couplings. The 
$\mu_i^*\lambda_{i\scriptscriptstyle 1\!1}$ combination contributes to 
electron EDM while the $\mu_i^*\lambda_{i\scriptscriptstyle 22}$ 
combination contributes to that of the muon. As we have no explicit 
numerical results to show at the moment, we refrain from showing any details
here.

There is in fact a second class of 1-loop diagrams contributing to the
quark EDMs. These are diagrams with quarks and scalars in the loop, and
hence superpartners of the charginolike and neutralinolike diagrams discussed
above. The basic formulae are also given in Ref.\cite{as6}. The R-parity
conserving analog of the class of diagrams has no significance, due to
the unavoidable small Yukawa couplings involved. With the latter replaced
by flavor-changing $\lambda^{\!\prime}$-couplings. We can have a $t$ quark
loop contributing to neutron EDM, for example. For the case of the
charged leptons, the two classes of superpartner diagrams merges into one.
But then, all scalars has to be included. The assumption hidden, in our
quark EDM formula above, that only the (two) superpartner sfermions
have a significant role to play does not stand any more. We have 
finished a $\mu \to \ e \gamma$ study, from which the charged lepton EDM 
formula could be extracted without too much effort\cite{as7}. Interested
readers may check the reference to get an idea, or just tune in for our
future publications\cite{as11}.

\newpage


\begin{table}[p]
\vspace*{.2in}
\caption{Numerical 1-loop neutron EDM results from SUSY without R parity, 
for four illustrative cases.  All EDM numbers are in $e\,\mbox{cm}$. 
Note that the quark EDM numbers are direct output from the numerical program
applying our quark dipole formulae; while the neutron EDM numbers are from
the valence quark model formula. 
R-parity violating parameters not given are taken as essentially 
zero. All parameters are taken real except those with complex phases explicitly
listed in each case, where the real number(s) listed then give the magnitude(s).
Parameter $A$ here means a common $A_u$ and $A_d$. Only 
$M_{\scriptscriptstyle 2}$ is shown for the gaugino masses; the others are
fixed by the unification relationship. Explicitly, we use 
$M_{\scriptscriptstyle 1}=0.5\,M_{\scriptscriptstyle 2}$ and
$M_{\scriptscriptstyle 3}=3.5\,M_{\scriptscriptstyle 2}$. The first column under 
``EDM Results" gives the couplings of the loop vertices involved. A $g$ indicates
either one of the electroweak gauge couplings, while a $\lambda^{\!\prime}$ 
coupling means one with the appropriate admissible flavor indices. In
the explicit results of the four cases, the latter is always a
$\lambda^{\!\prime}_{\scriptscriptstyle 31\!1}$. The second column gives
the reference Feynman diagram figures, when available. The third column indicates
whether the particular contribution involves a $LR$ squark mixing. In the case
that the mixing is involved and a R-parity violating (RPV) one is involved in
generating a RPV EDM contribution, it is marked with ``RPV". }

\vspace*{.2in}
\begin{tabular}{|c|c|c|c|c|c|c|}
\multicolumn{6}{|c}{\hspace*{1.4in}\framebox[3in][c]{\underline{\bf Choice of Parameters}}	} &	\\
\multicolumn{6}{|r}{$\tilde{m}_{\scriptscriptstyle Q}=300\,\mbox{GeV}$,
$\tilde{m}_{\scriptscriptstyle u}= \tilde{m}_{\scriptscriptstyle d}=200\,\mbox{GeV}$, 
$A=M_{\scriptscriptstyle 2}=300\,\mbox{GeV}$,
 $\mu_{\scriptscriptstyle 0}=-300\,\mbox{GeV}$}	 & \\
\hline
\multicolumn{3}{|c|}{$\tan\!\zb$} 				
& $3$		& $3$		& $3$		& $50$\\
\multicolumn{3}{|c|}{$\mu_{\scriptscriptstyle 3}$}	
& $1\cdot 10^{-3}\,\mbox{GeV}$	& $1\,\mbox{GeV}$		& $1\,\mbox{GeV}$		& $5\cdot 10^{-3}\,\mbox{GeV}$	\\
\multicolumn{3}{|c|}{$\lambda^{\!\prime}_{\scriptscriptstyle 31\!1}$}	
&  $0.05 $	& $0.05 $	& $0.05 $	& $0.05 $	\\
\multicolumn{3}{|c|}{(complex phases)} 
& $\lambda^{\!\prime}_{\scriptscriptstyle 31\!1}$($\pi/4$) 
& $\lambda^{\!\prime}_{\scriptscriptstyle 31\!1}$($\pi/4$)
& $\mu_{\scriptscriptstyle 0}$($0.5^o$), $A$($10^o$)	
& $\mu_{\scriptscriptstyle 0}$($0.02^o$),  $\mu_{\scriptscriptstyle 3}$($-\pi/4$)
\\
\hline
\multicolumn{6}{|c}{\hspace*{1.4in}\framebox[3in][c]{\underline{\bf EDM RESULTS :-}} }
&	\\
couplings										& Fig.	& $LR$-mixing	
& 		{\bf	Case A}	 & {\bf Case B	} & {\bf Case C}	& {\bf Case D} 	
\\
\hline\hline
\multicolumn{3}{l|}{\framebox[1.3in][c]{\underline{\bf \boldmath $d$ quark EDM}} } & & & &	\\
\multicolumn{3}{|l|}{\underline{ Gluino loop} : -} & & & &	\\
$\za_{\!\scriptscriptstyle s}$						& 1		& RPV
& $8.8\cdot 10^{-28}$	& $8.8\cdot 10^{-25}$	& -$3.9\cdot 10^{-26}$	& -$6.7\cdot 10^{-29}$ 	\\
\hline
\multicolumn{3}{|l|}{\underline{ Neutralino-like loop} : -} & & & & 	\\
$g^2$											& 1		& RPV
& -$1.9\cdot 10^{-29}$	& -$1.9\cdot 10^{-26}$	& $8.3\cdot 10^{-28}$	& $2.7\cdot 10^{-30}$ 
\\
$g\cdot y_{\!\scriptscriptstyle d}$						& 2		& no			
	& $\sim 0$	& $\sim 0$		& -$1.6\cdot 10^{-27}$	& -$1.2\cdot 10^{-27}$
\\
$g\cdot \lambda^{\!\prime}_{i{\scriptscriptstyle 1\!1}}$		& 3		& no
& -$1.0\cdot 10^{-28}$	& -$1.0\cdot 10^{-25}$	& $1.1\cdot 10^{-27}$	& $1.1\cdot 10^{-27}$ 
\\
$y_{\!\scriptscriptstyle d}^2$						& 4		& RPV
& $9.7\cdot 10^{-37}$	& $9.7\cdot 10^{-34}$	& -$3.9\cdot 10^{-35}$	& -$2.6\cdot 10^{-33}$
\\
$y_{\!\scriptscriptstyle d}\cdot\lambda^{\!\prime}_{ijk}$
												& 4		& yes
& -$1.7\cdot 10^{-36}$	& -$1.7\cdot 10^{-33}$	& $8.5\cdot 10^{-35}$	& $2.5\cdot 10^{-33}$
\\
two $\lambda^{\!\prime}_{ijk}$						& 4		& yes
& -$2.1\cdot 10^{-39}$	& -$3.4\cdot 10^{-34}$	& $9.0\cdot 10^{-35}$	& -$8.6\cdot 10^{-37}$
\\	
\hline
\multicolumn{3}{|l|}{\underline{ Chargino-like loop} : -}	 & & & & \\
$g\cdot y_{\!\scriptscriptstyle d}$						& 5		& no			
	& $\sim 0$ 	& $0$		& $2.5\cdot 10^{-26}$	& $1.7\cdot 10^{-26}$ 
\\
$g\cdot \lambda^{\!\prime}_{i{\scriptscriptstyle 1\!1}}$		& 6		& no
& $2.1\cdot 10^{-27}$	& $2.1\cdot 10^{-24}$	& -$1.3\cdot 10^{-26}$	& -$1.7\cdot 10^{-26}$ 
\\
$y_{\!\scriptscriptstyle u}\cdot y_{\!\scriptscriptstyle d}$	& 7		& yes			
	& $\sim 0$ 	& $0$		& -$2.7\cdot 10^{-34}$	& -$8.0\cdot 10^{-36}$
\\
$y_{\!\scriptscriptstyle u}\cdot\lambda^{\!\prime}_{ijk}$
											& 7		& yes
& -$2.1\cdot 10^{-37}$	& -$2.1\cdot 10^{-33}$	& $3.8\cdot 10^{-34}$	& $8.3\cdot 10^{-36}$
\\
\hline\hline
\multicolumn{3}{l|}{\framebox[1.3in][c]{\underline{\bf \boldmath $u$ quark EDM}} } & & & &	\\
\multicolumn{3}{|l|}{\underline{ Gluino loop} : -} & & & &	\\
$\za_{\!\scriptscriptstyle s}$						& 1		& yes
	& $0$		& $0$ 	& $4.5\cdot 10^{-26}$ & -$1.8\cdot 10^{-30}$
\\
\hline
\multicolumn{3}{|l|}{\underline{ Neutralino-like loop} : -} & & & & 	\\
$g^2$											& 1		& yes
	& $\sim 0$ 	& $0$		& $2.6\cdot 10^{-27}$	& -$1.4\cdot 10^{-31}$
\\
$g\cdot y_{\!\scriptscriptstyle u}$					& 2		& no	
	& $\sim 0$ 	& $0$		& $2.1\cdot 10^{-28}$	& $5.3\cdot 10^{-31}$
\\
$y_{\!\scriptscriptstyle u}^2$						& 4		& yes
	& $\sim 0$ 	& $0$		& $1.3\cdot 10^{-37}$	& $4.0\cdot 10^{-41}$
\\
\hline
\multicolumn{3}{|l|}{\underline{ Chargino-like loop} : -}	& & & & \\
$g\cdot y_{\!\scriptscriptstyle u}$						& 5		& no			
	& $\sim 0$ 	& $\sim 0$ 	& -$1.3\cdot 10^{-27}$	& -$3.2\cdot 10^{-30}$
\\
$y_{\!\scriptscriptstyle u}\cdot y_{\!\scriptscriptstyle d}$	& 7		& RPV			
& -$7.6\cdot 10^{-36}$	& -$7.6\cdot 10^{-33}$	& $3.2\cdot 10^{-34}$	& $6.4\cdot 10^{-34}$
\\
$y_{\!\scriptscriptstyle u}\cdot\lambda^{\!\prime}_{ijk}$
											& 7		& RPV
& $9.7\cdot 10^{-36}$	& $9.7\cdot 10^{-33}$	& $-5.1\cdot 10^{-34}$	& -$6.4\cdot 10^{-34}$
\\
\hline
\multicolumn{3}{l|}{\framebox[1.3in][c]{\underline{\bf Neutron EDM}} } & & & &	\\
\multicolumn{3}{|l|}{\underline{ from Gluino loop} : } 
& $1.8\cdot 10^{-27}$	& $1.8\cdot 10^{-24}$	& -$1.0\cdot 10^{-25}$	& -$1.4\cdot 10^{-28}$
\\
\multicolumn{3}{|l|}{\underline{ from Chargino-like loop} : } 
& $4.3\cdot 10^{-27}$	& $4.3\cdot 10^{-24}$	& $2.5\cdot 10^{-26}$	& $2.4\cdot 10^{-28}$	
\\
\multicolumn{3}{|l|}{\underline{ from Neutralino-like loop} : } 
& -$2.9\cdot 10^{-28}$	& -$2.9\cdot 10^{-25}$	& -$8.6\cdot 10^{-28}$	& -$2.0\cdot 10^{-29}$
\\
\multicolumn{3}{|l|}{\underline{TOTAL} : }
& $5.8\cdot 10^{-27}$	& $5.8\cdot 10^{-24}$	& -$7.8\cdot 10^{-26}$	& $8.0\cdot 10^{-29}$
\\
\end{tabular}
\end{table}

\end{document}